\documentclass[review]{elsarticle}

\usepackage{lineno}
\modulolinenumbers[5]

\usepackage[margin=1in]{geometry}
\usepackage{tikz}
\usepackage{graphicx}
\usetikzlibrary{arrows}
\usepackage{caption, subcaption}
\usepackage{multirow}
\usepackage{booktabs}
\usepackage{lscape}
\usepackage{rotating}   
\usepackage{amsmath}
\usepackage{array}
\usepackage{enumerate}
\usepackage[siunitx]{circuitikz}

\tikzstyle{block} = [rectangle, draw, fill=white!20, 
    text width=6em, text centered, minimum height=5em]
\tikzstyle{line} = [draw, -latex, line width = 0.8pt]   

\begin{document}

\begin{frontmatter}

\title{Modelling Noise and Pulse Width Modulation Interference in \\ Indoor Visible Light Communication Channels}

\author{Daniel~G.~Holmes}
\author{Ling~Cheng\corref{cor1}}
\cortext[cor1]{Corresponding author}
\ead{ling.cheng@wits.ac.za}
\author{Mulundumina~Shimaponda-Nawa\corref{cor2}}
\author{Ayokunle~D.~Familua\corref{cor3}}
\address{School of Electrical and Information Engineering, University of the Witwatersrand, Private Bag 3, Wits, 2050, Johannesburg, South Africa}

\author{Adnan~M.~Abu-Mahfouz\corref{cor4}}
\address{Modelling and Digital Science, Council for Scientific and Industrial Research (CSIR), Pretoria, 0184, South Africa}

\begin{abstract}
Visible light communication (VLC) has the potential to supplement the growing demand for wireless connectivity. In order to realise the full potential of VLC, channel models are required Discrete channel models based on semi-hidden Markov models (Fritchman model) for indoor VLC using low data rate LEDs are presented. Each channel considered includes background noise and differing types of interference from fluorescent lights and pulse-width modulated (PWM) LEDs, which could be part of an indoor smart lighting system. Models were developed based on experimental error sequences from a VLC system using an on-off keying (OOK) modulation scheme. The error sequences were input into the Baum-Welch algorithm to determine the model parameters by expectation maximisation. Simulated error sequences generated by the models are compared to and, in most cases, perform better than simpler models with a single bit error rate. The models closely approximate the experimental errors sequences in terms of error distribution. The models performed better in channels where there is less interference. It was also found that periodic errors were introduced as a results of the PWM modulated smart lighting LEDs. These models have use for designing error control codes and simulating indoor VLC environments with different types of interference.
\end{abstract}


\end{frontmatter}

\section{Introduction}

The demand for wireless connectivity is growing at an increasing rate. However, the current radio frequency bandwidth is limited and expensive. Visible light communication (VLC) utilising light-emitting diodes (LEDs) has drawn increasing attention as it has the potential to be part of the next generation of wireless connectivity to help meet this demand \cite{Pathak2015}. The ubiquitous nature of LEDs provides a major advantage for VLC to be employed in the indoor environment for both illumination and wireless connectivity, as part of technologies such as Li-Fi \cite{Haas2015} or smart lighting systems \cite{Sevincer2013, Sharma2018}. The other potential applications for VLC include indoor localisation \cite{Sahin2015}, vehicle-to-vehicle (V2V) communication \cite{Takai2014} and underwater communication \cite{Shen2016}.

One of the challenges with indoor VLC is dealing with noise and interference. This comes from both natural light as well as indoor lighting, both modulated and unmodulated. The transition to LED technology for all indoor lighting is still taking place. Therefore, early VLC systems may be deployed in environments where lighting other than LEDs exist. Furthermore, the potential co-existence of VLC with smart lighting systems presents a unique challenge where non-transmitting lighting levels become more dynamic, such as with pulse-width modulated (PWM) LED lights.

In \cite{Moreira1995}, artificial light interference from fluorescent and incandescent lights were characterised and modelled using experimental measurements in optical wireless channels. It was found that fluorescent lights with their wider band have higher potential to degrade the channel. The authors also showed that other sources of light can potentially cause errors in transmission, particularly in narrowband applications with low data rates. The work in \cite{Kizilirmak2015} proposed a method to mitigate the interference caused by a second LED based luminary in an orthogonal frequency division multiplexing (OFDM) VLC system. The interference was caused during the pulse transitions of the PWM signal. This was despite having an FFT at the receiver which filtered out the DC component of the PWM signal. The overlapping PWM pulse transitions degraded the useful OFDM signals.

In order to mitigate these errors, a better understanding of the VLC channel errors is required. One way to achieve this is through channel modelling. Existing work include into VLC channel modelling include optical power \cite{Chun2012, Komine2004}, multipath \cite{Lee2011a, Lu2015}, and ray tracing \cite{Sarbazi2014, Miramirkhani2015, Miramirkhani2017} channel models. However, research into VLC channel models have been sporadic and few works derive their channel models from experimental data which contribute to channel models that are more practical. Furthermore, there has been little work done on modelling the effects of noise or PWM interference.

One approach in modelling channel errors is to use a discrete channel model which models the channel at a symbol leve \cite{Ranjan2015}. A semi-hidden Markov model (SHMM) proposed by Fritchman can be used for discrete channel modelling of a digital channel \cite{Fritchman1967}. The Fritchman model has been used in the past to provide a statistical distribution of errors in wireless channels, particularly channels with bursty errors \cite{Familua2017}. These models are also able to simulate error sequences with error patterns similar to real channels. In \cite{Familua2015}, the authors used a three-state Fritchman model to model an frequency shift keying, on-off keying (FSK-OOK) PLC and VLC integrated system. Results showed a distribution of errors introduced from both the VLC and PLC channels for first and second order Markov models.  In \cite{VanHeerden1992}, a frequency hopped very high frequency (VHF) model based on the Fritchman model was developed using real time measurements from different modulation schemes. Similar work is reported in \cite{Chouinard1989a} where the Fritchman model was used in modelling digital radio channels where the model parameter estimation was achieved using the gradient method.

This paper presents Fritchman models that have been developed for several low bit rate indoor VLC cases with differing types of noise and interference. One such source of interference are PWM modulated LEDs which can operate at frequencies of less than 1~kHz \cite{IEEE2015}. Not all VLC applications may require high frequency transmission, such as for indoor localisation. As such, the interest is in low bit rate transmission where the impact of interfering PWM signals could be more significant. The parameters of the channel models were found by expectation maximisation using the Baum-Welch algorithm \cite{Baum1970}. Error sequences obtained from transmissions run on a VLC test-bed were used as input data for the Baum-Welch algorithm to train the models. Once the parameters for the models were found, the models were used to simulate channel error sequences.  A statistical model for channel errors can help in the design of error control codes, interference avoidance and mitigation in future smart lighting systems, and as part of indoor channel VLC system software simulations. Additionally, the Fritchman model can potentially provide a better distribution of channel errors than a single bit error rate (BER) value.

The rest of the paper is organised as follows. Section~\ref{sec:channel-modelling} provides a theoretical background on the modelling techniques used. Section~\ref{sec:system-description} gives a description of the system that was used to collect the experimental data. Section~\ref{sec:experimental-procedure} explains the experimental procedure. Section~\ref{sec:results-discussion} presents the results along with discussions, followed by a conclusion in Section~\ref{sec:conclusion}.

\section{Channel Modelling}\label{sec:channel-modelling}

\subsection{Modelling Approach}

There are two approaches modelling a channel behaviour. The first is applying a signal level approach where the channel is modelled in terms of signal parameters. These are defined as waveform-level channel models and would typically include models for the signal-to-noise-ratio (SNR), signal-to-interference-plus-noise ratio (SINR), multipath, optical power and the Lambertian model. The other approach to channel modelling is using a statistical or probabilistic model, such as the one proposed for this research, which gives a statistical distribution of channel errors. This type of channel model is different to the other types of models because it provides a high level model of how errors are distributed in a channel. Thus, the model is considered discrete due to the fact that individual channel states are considered.

Modelling the errors on a channel using a single error probability is simple, yet it cannot describe more complex error distributions and patterns. The Fritchman model has multiple states and error probabilities. This means that the Fritchman model can potentially provide a better overall model for channel error distributions and patterns \cite{tranter2004principles}. Channel modelling for VLC in general has not been studied considerably. The application of the Fritchman model in modelling a VLC channel errors from noise as well as PWM interference is not evident in literature either. There have been several efforts to determine physical channel models where a signal approach has been taken \cite{ghassemlooy2017theory}. A statistical or probabilistic approach in determining channel error distributions have been applied in other communication fields.

Discrete channel models are different from waveform (signal level) models in that they abstract out the waveform signal. A waveform signal consists of a sampled combination of the transmitted signal as well as disturbances such as noise and interference, whereas a discrete channel model is in terms of symbols only. The noise and interference are not modelled directly, but rather the resultant combination of these with the signal is modelled. Discrete channel models are favourable because they are computationally more efficient when compared to waveform channel models. The model can be characterised in terms of a small set of parameters because the many physical aspects of the channel are abstracted out. An important part of the modelling process is determining these parameters. This can be accomplished by physical measurements on the actual channel.

\subsection{Fritchman Model}

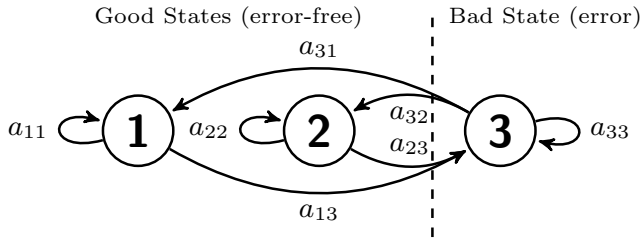
\begin{figure}[t!]
\centering
\resizebox{9cm}{!}{
\begin{tikzpicture}[->,>=stealth',shorten >=1pt,auto,node distance=2cm,
                    thick,main node/.style={circle,draw,font=\sffamily\Large\bfseries}]

  \node[main node] (1) {1};
  \node[main node] (2) [right of=1] {2};
  \node[main node] (3) [right of=2] {3};
  
  \draw[dashed, thick, -] (3.25,1.25) -- (3.25,-1.25);
  
  \node at (4.5,1.25) {\scriptsize{Bad State (error)}};
  \node at (1,1.25) {\scriptsize{Good States (error-free)}};

  \path[every node/.style={font=\sffamily\small}]
    (1) edge [bend right] node [below] {$a_{13}$} (3)
        edge [loop left] node {$a_{11}$} (1)
    (2) edge [bend right] node [above] {$a_{23}$} (3)
        edge [loop left] node {$a_{22}$} (2)
    (3) edge [bend right] node [above] {$a_{31}$} (1)
        edge [bend right] node [below] {$a_{32}$} (2)
        edge [loop right] node [right] {$a_{33}$} (3);
\end{tikzpicture}
}
\caption{A three-state Fritchman model with two good states and one bad state.}
\label{fig:3-state}
\end{figure}

For the case of binary channels, the Fritchman model framework partitions the channel state space into good and bad states. Fritchman defined $k$ good states representing error-free transmissions and $N - k$ bad states representing a transmission where an error always occurs. Fig.~\ref{fig:3-state} shows a three-state Fritchman model with two good states and one bad state. Each state has a set of transition probabilities which are used as part of the model. They form part of the state transition matrix:

\begin{equation}
	\mathbf{A} = \begin{bmatrix} 
	a_{11} & 0 & a_{13} \\ 
	0 & a_{22} & a_{23} \\ 
	a_{31} & a_{32} & a_{33}
	\end{bmatrix}
	\label{eqn:A}
\end{equation}

Note that for a Fitchman model, there are no transitions between good states, making the elements $a_{12}$ and $a_{21} $ zero. Another parameter used in the model is the error generation matrix, $ \mathbf{B} $. It describes the probability of generating an error within a specific state at a discrete point in time. Bearing in mind the fact that good states are error-free and bad states always produce an error, the error generation matrix for a three-state model is thus: 

\begin{equation}
	\mathbf{B} = \begin{bmatrix} 
	1 & 1 & 0 \\ 
	0 & 0 & 1  
	\end{bmatrix}
	\label{eqn:B}
\end{equation}

This describes how the model is semi-hidden. Even though one can observe whether the channel is in a good or a bad state, based on the error output, one cannot observe as to which of the good states produced the error-free transmission. The third parameter of the model is the initial state probability for any of the three states:

\begin{equation}
	\mathbf{\Pi} = \begin{bmatrix} 
	\pi_{1} & \pi_{2} & \pi_{3}
	\end{bmatrix}
	\label{eqn:pi}
\end{equation}

\subsection{Baum-Welch Algorithm}

The Baum-Welch algorithm is a robust method for fitting a SHMM. Using this algorithm, the parameters for the channel model, $ \Gamma = (\mathbf{A}, \mathbf{B}, \mathbf{\Pi}) $, can be estimated. This is an iterative algorithm that uses either a measured or simulated error sequence, $ {\overline{O} = \lbrace O_{1}, O_{2}, ... O_{t}, ... O_{T}\rbrace} $, to converge to the maximum likelihood estimator for the model parameters that maximizes $\text{Pr}(\overline{O}|\Gamma)$. The number of iterations depends on the desired level of accuracy for the model \cite{tranter2004principles}.

\subsection{Model Scenarios}\label{sec:scenarios}

Some examples of light that can add noise and interference to VLC channels include light from the sun and other indoor lighting such as fluorescent and incandescent lights. Another source of interference are other LED lights. These include LEDs that are not transmitting data, but are instead PWM modulated to support dimming of the smart lighting. Models were developed for two different indoor channels with different noise and interference present, including:
 
\begin{enumerate}[I]
	\item Background noise from sunlight passing through windows and from fluorescent indoor lighting
	\item The same background noise as case~I as well as an interfering PWM modulated LED. This case includes three sub-cases for different PWM duty cycles or dimming levels, namely 25\%, 50\% and 75\%.
\end{enumerate} 

Experiments for each of these cases took place in a lab environment that includes a number of windows and fluorescent lights.

\section{System Description}\label{sec:system-description}

\subsection{Hardware}

\begin{figure}[t]
	\centering
	\begin{subfigure}[b]{0.5\textwidth}
	\centering
		\includegraphics[scale=0.05,height=8cm]{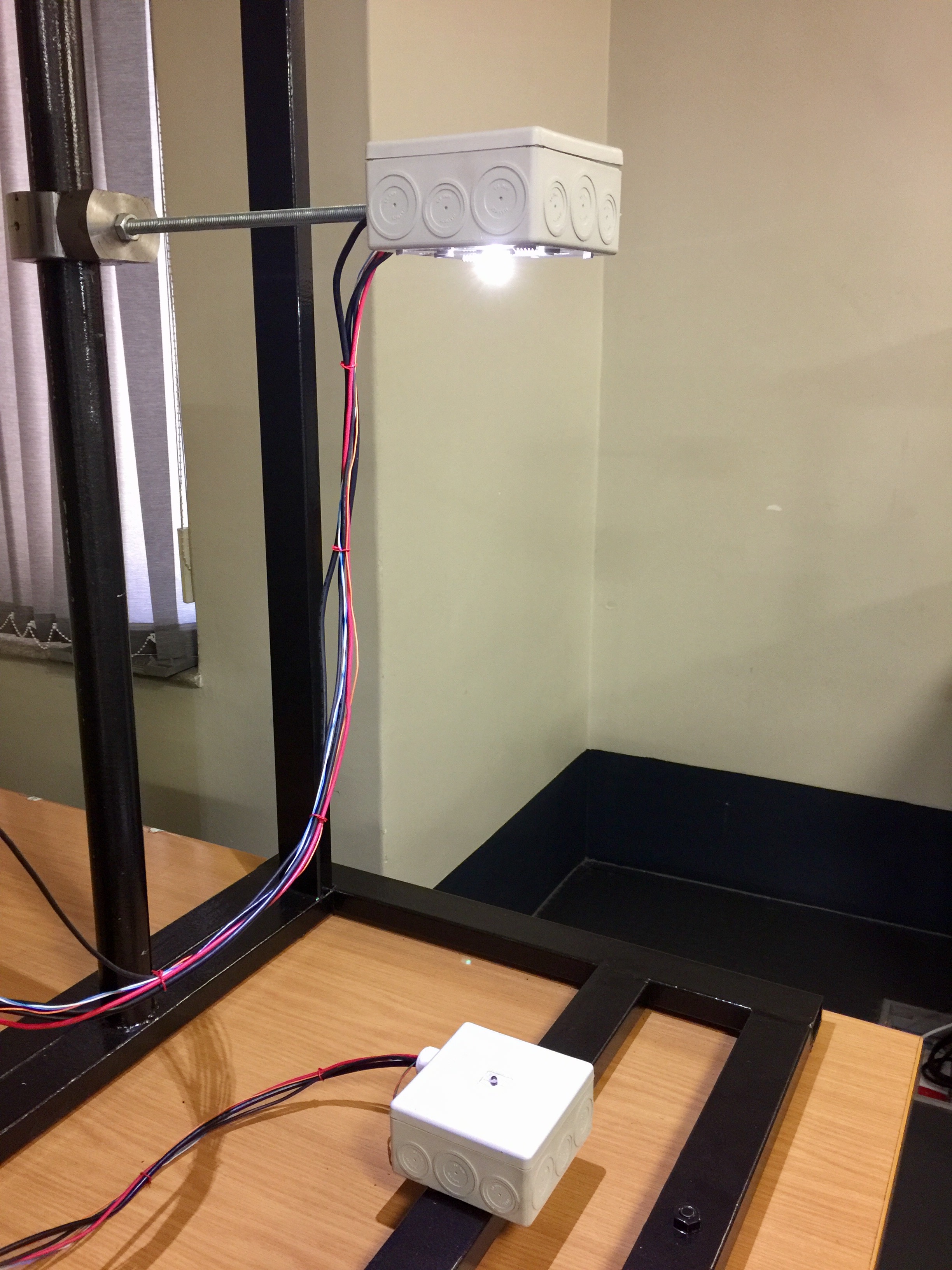}
		\caption{Setup with adjustable height.}
	\end{subfigure}%
	~
	\begin{subfigure}[b]{0.35\textwidth}
	\centering
		\includegraphics[scale=0.035,height=3.6cm]{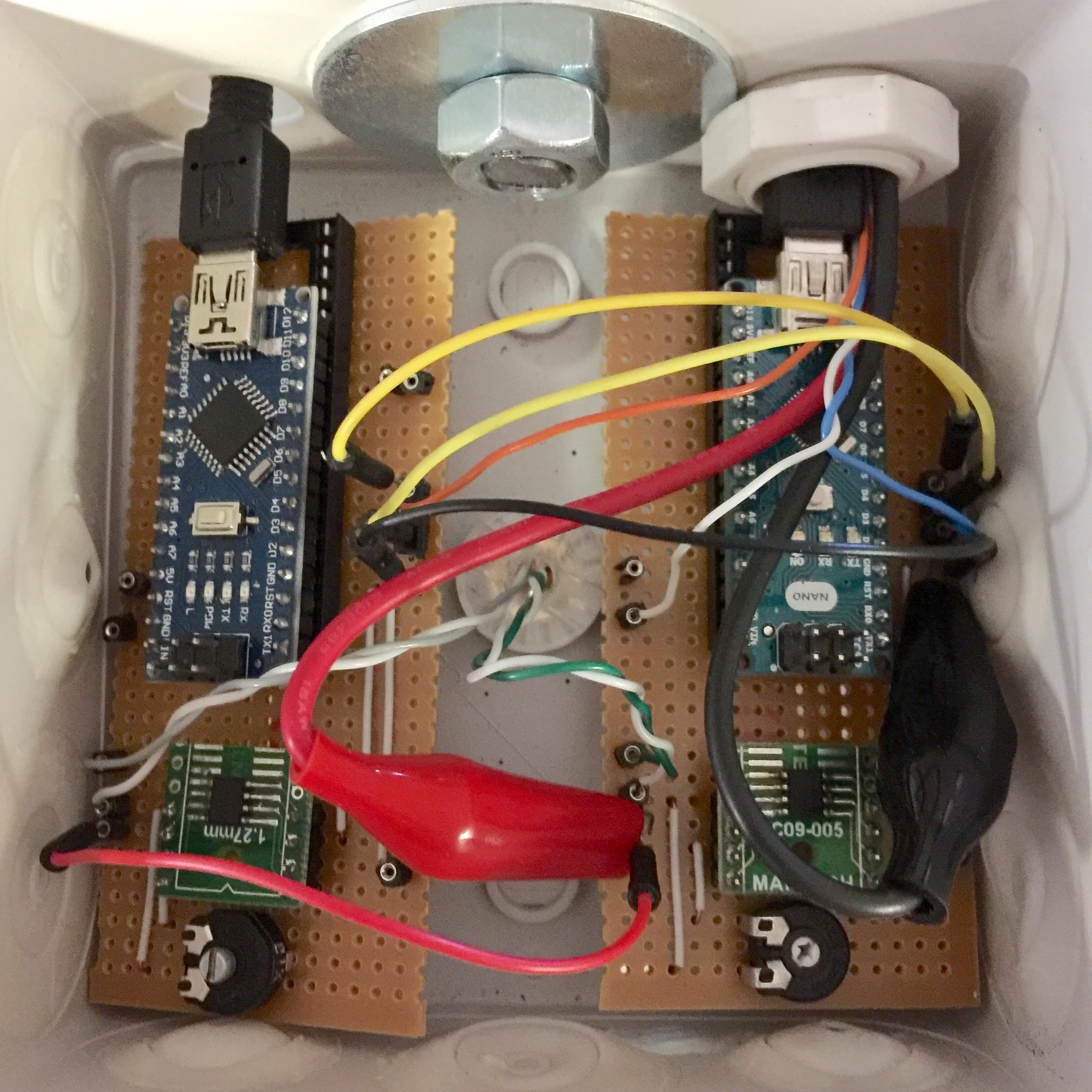}
		\caption{Transmitter circuit.}
		\includegraphics[scale=0.035,height=3.6cm]{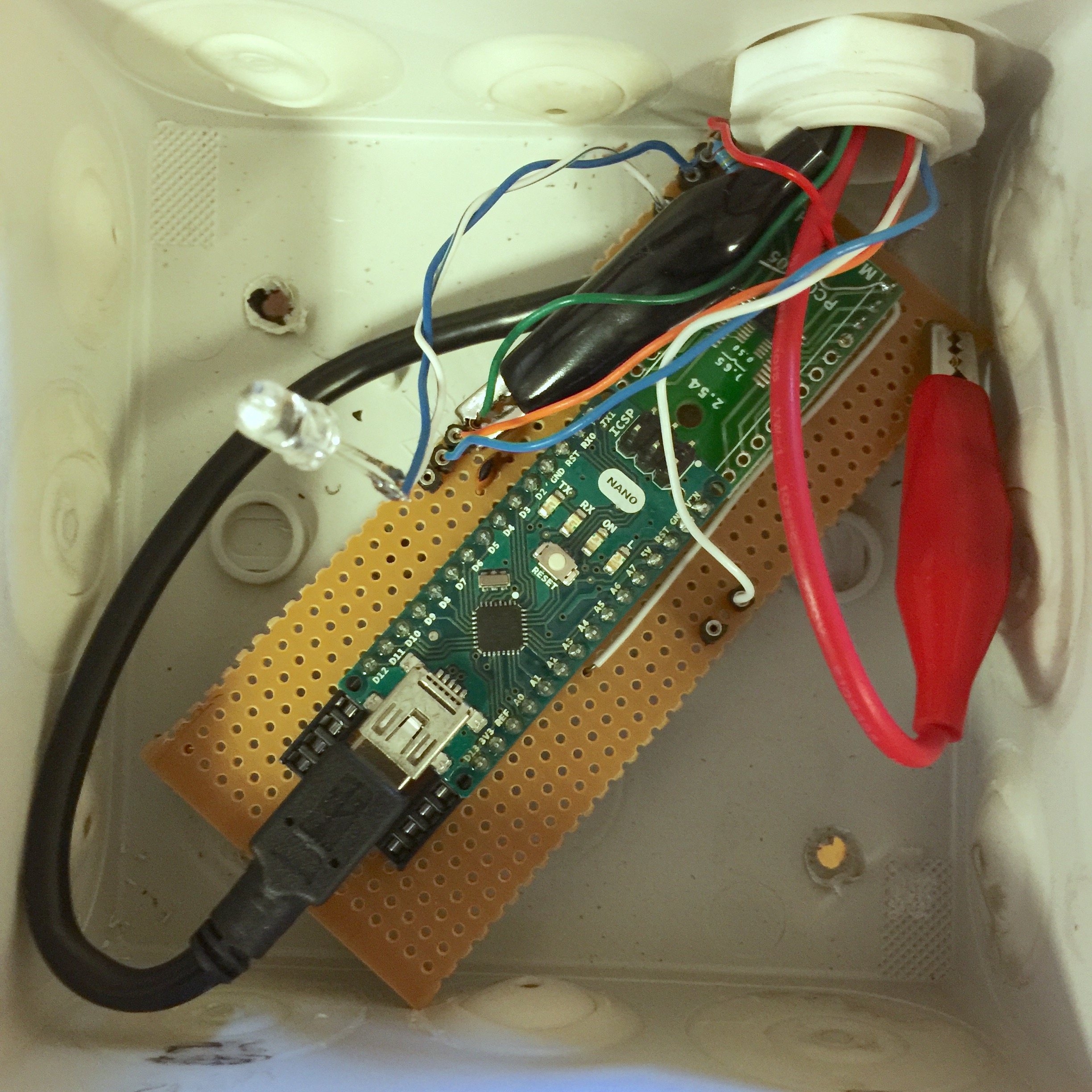}
		\caption{Receiver circuit.}
	\end{subfigure}
\caption{Pictures of the full test-bed and hardware.}
\label{fig:hardware}
\end{figure}

\begin{figure}[t]
\centering
\includegraphics[scale=0.35]{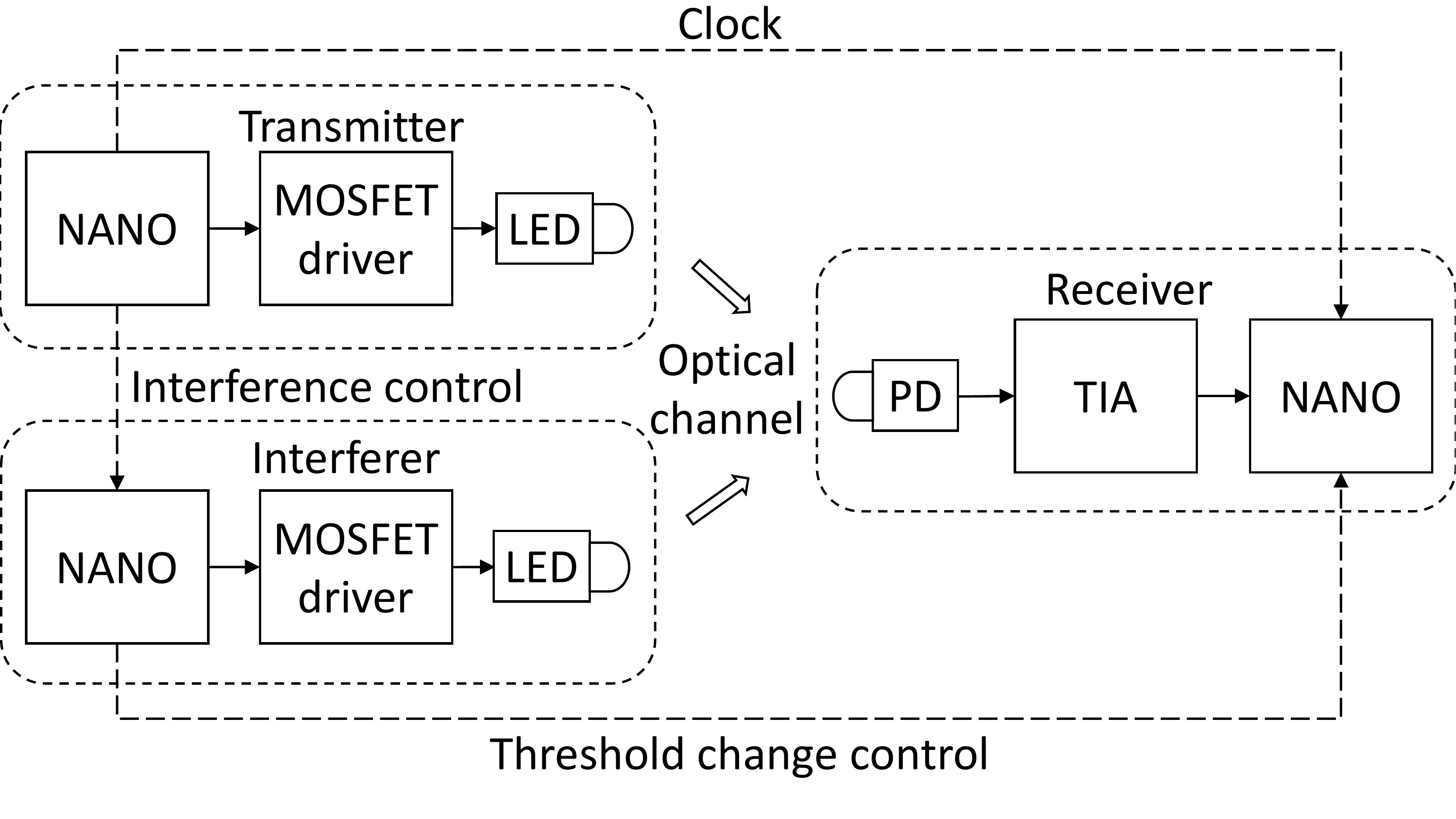}
\caption{Block diagram of VLC system used to obtain error sequences. PD, photodiode; TIA, transimpedance amplifier.}
\label{fig:system}
\end{figure}

Fig.~\ref{fig:hardware} shows a photograph of the experimental setup with the transmitter and receiver boxes in an indoor environment. Additional information about the LED used for the experiments is found in Table~\ref{tab:parameters}. Fig.~\ref{fig:system} shows a detailed block diagram of the system, which includes transmitter, interferer and receiver modules.

\begin{table}
\centering
\caption{Experimental hardware parameters.}
\begin{tabular}{ll}
\hline 
\hline
LED make & Cree$^{\textregistered}$ XLamp$^{\textregistered}$ XP-G \\ 
Transmitter viewing angle & $125^{\circ}$ \\ 
Separation distance & 1 m \\ 
Photodiode make & OSRAM SFH 213 \\ 
Receiver half angle & $\pm 10^{\circ}$ \\ 
\hline 
\end{tabular} 
\label{tab:parameters}
\end{table}

The transmitter's Arduino Nano outputs an OOK signal to the MOSFET circuit which then drives the white LED. The OOK signal is received at the photodiode and then converted from a current signal to a voltage signal by means of a transimpedance amplifier (TIA). The analog-to-digital converter (ADC) at the receiver's Nano samples the signal and determines whether the received bit is a 1 or a 0 based on a decision threshold.  Due to the low bit rate transmission, the receiver does not include a DC filter. Communication between the transmitter and receiver is synchronised by a clock line. This clock line is necessary so that no synchronisation errors occur. The only errors we are interested in should come as a result of the channel. The interferer transmits a signal in the same manner as the transmitter. The transmitter module controls when the interferer should be activated as part of the experimental procedure.

\subsection{Decision Threshold}

\begin{figure}[t]
\centering
\includegraphics[scale=1]{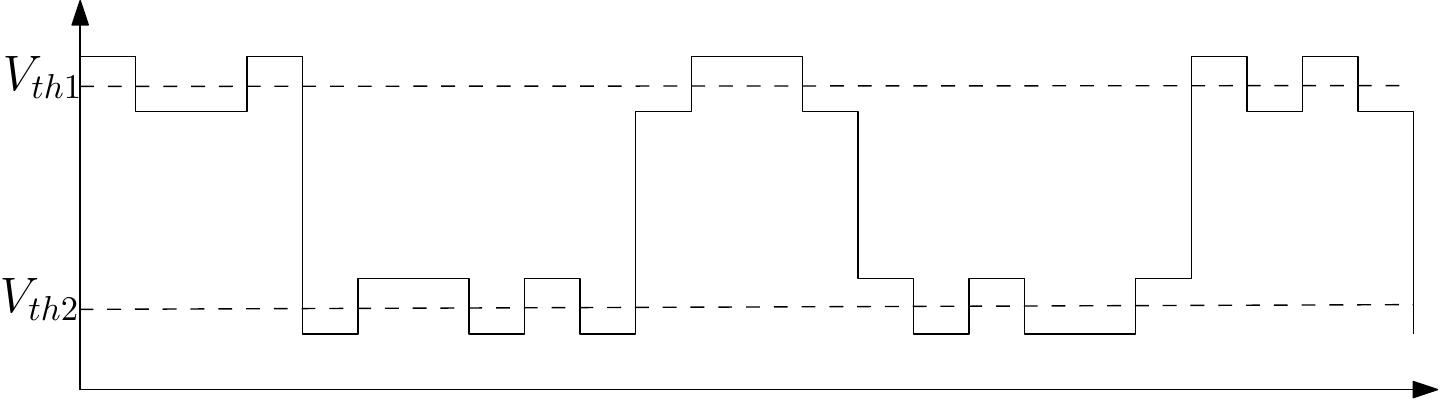}
\caption{Voltage waveform with random bit sequence, PWM interference, and background noise DC offset measured at the receiver, showing two decision thresholds.}
\label{fig:case2-vth}
\end{figure}

At the receiver, the ADC threshold value is different for each experimental case. For case~I, there is a single threshold value. Fig.~\ref{fig:case2-vth} shows an example of the two thresholds used for case~II at the receiver. $V_{th1}$ is used when the interfering PWM signal is on and $V_{th2}$ for when it is off. This is because when the PWM signal is on, it simply adds a DC offset to the received signal. Note that the PWM signal has an additional DC offset as a result of background noise. The interferer informs the receiver when it needs to change its threshold value while a transmission is taking place. The justification for this approach is that a PWM signal is deterministic and it is possible to change threshold without a connection between the interferer and receiver. However, the implementation of this is beyond the scope of the work presented in this paper. These approaches result in an optimal threshold detections for cases~I and II.

Threshold values are determined at the beginning of each transmission by sending pilot bit sequences. The pilot bits are sent once for case~I. For case~II, the pilot bits are sent twice, once for when the PWM signal is on, and again for when the PWM signal is off. The same pilot bits are also sent at the end of a transmission. 

\section{Experimental Procedure}\label{sec:experimental-procedure}

For each case, 100~000 pseudo-random bits were transmitted at a rate of 6.25~kbit/s in 10~000 bit chunks. The interfering PWM signal has a frequency of 600~Hz with duty cycle values as per the case~II description (see Section \ref{sec:scenarios}). Transmissions were run for all three of these case~II duty cycles. Transmission distance was fixed at 1~m. In order to get a variety of modelling results, a number of transmissions took place for each case within a range of SNR in case~I, or SINR in case~II. This was done by varying the optical power of the transmitter.

The SNR and SINR range was chosen based on the number of errors in the error sequence which was between approximately 1\% and 10\%. The received binary sequence was compared with the known transmitted sequence to obtain an error sequence. A $1$ in the error sequence denotes an error and a $0$ denotes no error. This was then used as the input to the Baum-Welch algorithm.

Along with the error sequences collected for each case, initial values for the model parameters were used as the input for the Baum-Welch algorithm, which include:

\begin{equation}
	\mathbf{A} = \begin{bmatrix} 
	0.9 & 0 & 0.1 \\ 
	0 & 0.8 & 0.2 \\ 
	0.1 & 0.7 & 0.2
	\end{bmatrix}
\end{equation}

\begin{equation}
	\mathbf{B} = \begin{bmatrix} 
	1 & 1 & 0 \\ 
	0 & 0 & 1  
	\end{bmatrix}
\end{equation}

\begin{equation}
	\mathbf{\Pi} = \begin{bmatrix} 
	0.4 & 0.4 & 0.2
	\end{bmatrix}
\end{equation}

For case~I, the SNR was calculated using:

\begin{equation}
\label{eqn:snr}
	\text{SNR} = \dfrac{\sigma_{S}^{2}}{\sigma_{B}^{2}}
\end{equation}

The $ \sigma_{B}^{2} $ term includes the background, thermal and shot noises as well ($ \sigma_{shot}^{2} + \sigma_{thermal}^{2} $). This measurement was taken in single readings for each test. For case~II, the SINR was calculated using three measurements. A description of each is followed by the relevant equation below. The first and second measurements are used to get the signal component while the third measurement is used to get the noise and interference components.

\begin{enumerate}
	\item PWM signal is active:
	\begin{equation}
		\label{eqn:sigma1}
		\sigma_{1}^{2} = \sigma_{B}^{2} + \sigma_{PWM}^{2}
	\end{equation}
	
	\item PWM signal is active and the source is transmitting a random signal:
	\begin{equation}
		\label{eqn:sigma2}
		\sigma_{2}^{2} = \sigma_{B}^2 + \sigma_{PWM}^{2} + \sigma_{S}^{2}
	\end{equation}

	\item PWM signal is active with measurements $ X_{i} $ grouped based on whether the PWM signal is high or low.

	\begin{equation}
	\label{eqn:sigma3}
		\sigma_{3}^{2} \Rightarrow \lbrace X_{i} \rbrace : T_{low}
	\end{equation}

	\begin{equation}
		\sigma_{4}^{2} \Rightarrow \lbrace X_{i} \rbrace : T_{high}
	\end{equation}

	\begin{equation}
		\dfrac{\sigma_{3}^{2} + \sigma_{4}^{2}}{2}  = \overline{\sigma}_{PWM}^{2} + \sigma_{B}^{2}
	\end{equation}
\end{enumerate}

Using the above measurements, the final SINR is then calculated by:

\begin{equation}
	\label{eqn:sigma4}
	\text{SINR} = \dfrac{2(\sigma_{2}^{2} - \sigma_{1}^{2})}{\sigma_{3}^{2} + \sigma_{4}^{2}}
		\end{equation}

\section{Results and Discussion}\label{sec:results-discussion}

\subsection{Model Derivation and Comparison}

The Baum-Welch algorithm was used to generate three-state Fritchman models. The resultant state transition probabilities of the models were used to generate new error sequences of 100~000 bits for the different channels at each SNR/SINR. In order to determine how well the models describe the channel, a comparison is made between a single BER value and the Fritchman model with its multiple states. Independent and identically distributed (IID) error sequences were generated using each sequence error probability, $ P_{e} $, obtained from the experimental error sequence. The measured, modelled and IID error sequences were then compared using error-free run distribution (EFRD) plots, expressed as $ \text{Pr}(0^m|1) $. This is described as the probability of transitioning to $ m $ or more consecutive error-free states following the occurrence of an error. The EFRD from the measured sequence is compared to the modelled and IID EFRD using the chi-squared ($\chi^2$) test and mean-squared error (MSE) to determine goodness of fit, as these are widely used for determining how well curves match. The log-likelihood function has also been used to determine the Baum-Welch algorithm convergence, as well as to determine how well the model fits the error sequence used to train the model.

\begin{figure}[bt]
	\centering
	\includegraphics[scale=0.4]{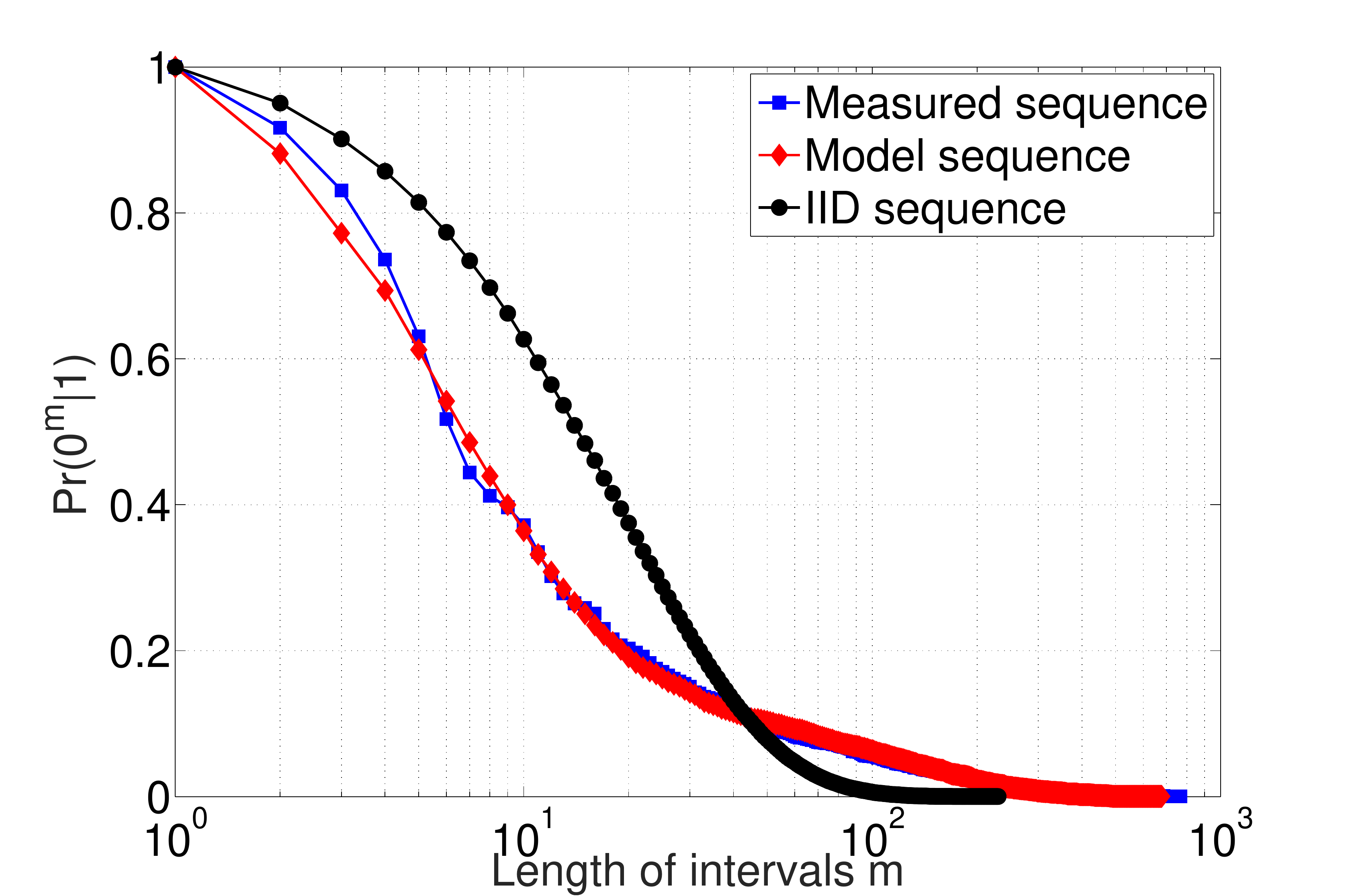}
        \caption{Channel with background noise from indoor lighting and windows at 6.66~dB SNR case~I.}
    \label{fig:case1}
\end{figure}

\begin{sidewaystable*}
  \centering
  \caption{Modelling Results for case~I - direct line-of-sight}
    \begin{tabular}{ccccccccccccc}
    \toprule
    \toprule
    \multirow{2}[4]{*}{SNR (dB)} & \multirow{2}[4]{*}{$ P_{e} $} & \multicolumn{7}{c}{Transition probabilities}          & \multicolumn{2}{c}{MSE} & \multicolumn{2}{c}{$ \chi^2 $} \\
\cmidrule{3-9} \cmidrule(lr){10-11} \cmidrule(lr){12-13}         &       & $ a_{11} $ & $ a_{13} $ & $ a_{22} $ & $ a_{23} $ & $ a_{31} $ & $ a_{32} $ & $ a_{33} $ & IID   & Model & IID   & Model \\
    \midrule
    7.68  & 0.0067 & 0.9960 & 0.0040 & 0.9724 & 0.0276 & 0.5338 & 0.4320 & 0.0342 & 0.00283 & 0.00087 & 22.67724 & 8.76799 \\
    6.95  & 0.0238 & 0.9916 & 0.0084 & 0.9181 & 0.0819 & 0.2843 & 0.6067 & 0.1090 & 0.00424 & 0.00023 & 13.70263 & 0.72130 \\
    6.66  & 0.0502 & 0.9895 & 0.0105 & 0.8614 & 0.1386 & 0.1481 & 0.6712 & 0.1807 & 0.00576 & 0.00007 & 9.33804 & 0.12776 \\
    6.44  & 0.0363 & 0.9848 & 0.0152 & 0.8862 & 0.1138 & 0.3317 & 0.5483 & 0.1200 & 0.00270 & 0.00011 & 6.80294 & 0.37627 \\
    5.91  & 0.0449 & 0.9700 & 0.0300 & 0.9196 & 0.0804 & 0.4766 & 0.4321 & 0.0913 & 0.00040 & 0.00017 & 0.82839 & 0.20040 \\
    5.60  & 0.0480 & 0.9729 & 0.0271 & 0.8293 & 0.1707 & 0.4832 & 0.3926 & 0.1242 & 0.00165 & 0.00010 & 2.67459 & 0.12581 \\
    5.48  & 0.0840 & 0.9556 & 0.0444 & 0.8703 & 0.1297 & 0.3005 & 0.5473 & 0.1522 & 0.00056 & 0.00012 & 0.78514 & 0.10873 \\
    4.73  & 0.0905 & 0.9426 & 0.0574 & 0.8970 & 0.1030 & 0.2366 & 0.6135 & 0.1499 & 0.00037 & 0.00023 & 0.32733 & 0.18721 \\
    4.50  & 0.1039 & 0.9226 & 0.0774 & 0.8918 & 0.1082 & 0.1861 & 0.6724 & 0.1415 & 0.00020 & 0.00014 & 0.11204 & 0.06155 \\
    4.28  & 0.1281 & 0.9345 & 0.0655 & 0.8373 & 0.1627 & 0.2078 & 0.5917 & 0.2005 & 0.00041 & 0.00013 & 0.42408 & 0.30211 \\
    \bottomrule
    \bottomrule
    \end{tabular}%
  \label{tab:case1-results}%
\end{sidewaystable*}%

\subsection{Case I}

Table \ref{tab:case1-results} shows the modelling results for case~I along with comparisons of the models and IID $\chi^2$ and MSE values. Each of the models for this case have lower MSE and $\chi^2$ values compared to the IID. This indicates that the Fritchman model provides a more accurate way to model errors in VLC channel with background noise, compared to a single BER value. Fig. \ref{fig:case1} shows one of the EFRD comparisons where the model follows the distribution of the measured sequence. The main difference between the IID and and the measured sequence is that the IID value gives an error distribution based on the BER and does not take into consideration the memory of the channel. Although the percentage of errors of the measured and IDD sequences are the same, the errors patterns are different. The Fritchman model takes into consideration the memory of the channel.

\subsection{Case II}

\subsubsection{Scenario Results}

Tables \ref{tab:case2-25-results}, \ref{tab:case2-50-results} and \ref{tab:case2-75-results} show the case~II modelling results for the 25\%, 50\% and 75\% duty cycles respectively. The first half of the tables show that the models generated for case~II have $ a_{33} = 0$. This represents the probability of having two or more consecutive errors in the error sequence, forming error clusters. This means there are no error clusters. The $\chi^2$ and MSE values show that the models do not always provide a better fit than the IID sequences in tests for case~II. The error sequence produced by the model is unable to follow the measured error sequence with significant improvement compared to the IID error sequence. It is evident that Fritchman models are better suited to channels that have error clusters.

At higher SINR tests, the majority of errors appear to be caused by the interfering PWM signal. Between about 5~dB and -10~dB, the tests show $P_e$ values that are consistently around 0.05. This is also around the region where $ a_{33} = 0 $. Thus, errors caused within this region are almost exclusively as a results of the interfering PWM signal. As the SINR decreases, errors are introduced from background noise as well, which increases the $P_e$ and gives $ a_{33} > 0 $.

\begin{sidewaystable*}
  \centering
  \caption{Modelling Results for case~II - 25\% Duty Cycle Interference}
    \begin{tabular}{ccccccccccccc}
    \toprule
    \toprule
    \multirow{2}[4]{*}{SINR (dB)} & \multirow{2}[4]{*}{$ P_{e} $} & \multicolumn{7}{c}{Transition probabilities}          & \multicolumn{2}{c}{MSE} & \multicolumn{2}{c}{$ \chi^2 $} \\
\cmidrule{3-9} \cmidrule(lr){10-11} \cmidrule(lr){12-13}         &       & $ a_{11} $ & $ a_{13} $ & $ a_{22} $ & $ a_{23} $ & $ a_{31} $ & $ a_{32} $ & $ a_{33} $ & IID   & Model & IID   & Model \\
    \midrule
    5.57  & 0.0495 & 0.9481 & 0.0519 & 0.9477 & 0.0523 & 0.4014 & 0.5986 & 0.0000 & 4.24E-04 & 4.58E-04 & 0.9953 & 0.4541 \\
    -1.02 & 0.0501 & 0.9480 & 0.0520 & 0.9467 & 0.0533 & 0.3931 & 0.6069 & 0.0000 & 3.64E-04 & 3.35E-04 & 0.6167 & 0.3674 \\
    -2.04 & 0.0494 & 0.9486 & 0.0514 & 0.9476 & 0.0524 & 0.4015 & 0.5985 & 0.0000 & 3.68E-04 & 3.20E-04 & 0.7938 & 0.2031 \\
    -5.35 & 0.0512 & 0.9478 & 0.0522 & 0.9448 & 0.0552 & 0.3863 & 0.6137 & 0.0000 & 2.65E-04 & 2.21E-04 & 0.5570 & 0.2362 \\
    -9.55 & 0.0527 & 0.9451 & 0.0549 & 0.9439 & 0.0561 & 0.3724 & 0.6276 & 0.0000 & 3.29E-04 & 2.80E-04 & 0.7966 & 0.4347 \\
    -13.84 & 0.0523 & 0.9466 & 0.0534 & 0.9437 & 0.0563 & 0.3781 & 0.6215 & 0.0004 & 3.50E-04 & 3.65E-04 & 0.5347 & 0.3648 \\
    -16.33 & 0.1012 & 0.9142 & 0.0858 & 0.8945 & 0.1055 & 0.1800 & 0.7154 & 0.1046 & 7.69E-05 & 9.83E-05 & 0.0291 & 0.0427 \\
    -16.92 & 0.1056 & 0.9152 & 0.0848 & 0.8888 & 0.1112 & 0.1723 & 0.7159 & 0.1118 & 7.12E-05 & 6.33E-05 & 0.0254 & 0.0352 \\
    -18.71 & 0.0739 & 0.9351 & 0.0649 & 0.9201 & 0.0799 & 0.2584 & 0.6836 & 0.0581 & 1.59E-04 & 1.31E-04 & 0.0556 & 0.0459 \\
    -19.24 & 0.0861 & 0.9308 & 0.0692 & 0.9062 & 0.0938 & 0.2215 & 0.6951 & 0.0834 & 1.01E-04 & 9.41E-05 & 0.0385 & 0.0458 \\
    \bottomrule
    \bottomrule
    \end{tabular}%
  \label{tab:case2-25-results}%
\end{sidewaystable*}%

\begin{sidewaystable*}
  \centering
  \caption{Modelling Results for case~II - 50\% Duty Cycle Interference}
    \begin{tabular}{ccccccccccccc}
    \toprule
    \toprule
    \multirow{2}[4]{*}{SINR (dB)} & \multirow{2}[4]{*}{$ P_{e} $} & \multicolumn{7}{c}{Transition probabilities}          & \multicolumn{2}{c}{MSE} & \multicolumn{2}{c}{$ \chi^2 $} \\
\cmidrule{3-9} \cmidrule(lr){10-11} \cmidrule(lr){12-13}         &       & $ a_{11} $ & $ a_{13} $ & $ a_{22} $ & $ a_{23} $ & $ a_{31} $ & $ a_{32} $ & $ a_{33} $ & IID   & Model & IID   & Model \\
    \midrule
    6.37  & 0.0486 & 0.9491 & 0.0509 & 0.9487 & 0.0513 & 0.3943 & 0.6057 & 0.0000 & 6.38E-04 & 6.01E-04 & 0.8494 & 0.4041 \\
    -0.58 & 0.0513 & 0.9461 & 0.0539 & 0.9458 & 0.0542 & 0.3744 & 0.6256 & 0.0000 & 6.74E-04 & 7.24E-04 & 1.0036 & 0.6290 \\
    -2.70 & 0.0575 & 0.9421 & 0.0579 & 0.9397 & 0.0603 & 0.3327 & 0.6426 & 0.0247 & 2.77E-04 & 2.26E-04 & 0.3368 & 0.1879 \\
    -6.20 & 0.0511 & 0.9466 & 0.0534 & 0.9459 & 0.0541 & 0.3754 & 0.6246 & 0.0000 & 6.01E-04 & 5.33E-04 & 0.7357 & 0.2321 \\
    -8.55 & 0.0518 & 0.9456 & 0.0544 & 0.9452 & 0.0548 & 0.3684 & 0.6316 & 0.0000 & 6.04E-04 & 5.45E-04 & 0.9099 & 0.4766 \\
    -12.03 & 0.0582 & 0.9404 & 0.0596 & 0.9388 & 0.0612 & 0.3291 & 0.6523 & 0.0186 & 3.09E-04 & 3.19E-04 & 0.4044 & 0.2140 \\
    -14.10 & 0.0586 & 0.9397 & 0.0603 & 0.9381 & 0.0619 & 0.3218 & 0.6641 & 0.0142 & 3.65E-04 & 3.62E-04 & 0.4813 & 0.2922 \\
    -14.64 & 0.0707 & 0.9295 & 0.0705 & 0.9263 & 0.0737 & 0.2658 & 0.6908 & 0.0434 & 1.77E-04 & 1.37E-04 & 0.1761 & 0.0717 \\
    -15.01 & 0.0912 & 0.9141 & 0.0859 & 0.9063 & 0.0937 & 0.1993 & 0.7163 & 0.0844 & 7.81E-05 & 7.53E-05 & 0.0484 & 0.0238 \\
    -16.32 & 0.0900 & 0.9168 & 0.0832 & 0.9067 & 0.0933 & 0.2022 & 0.7167 & 0.0811 & 6.82E-05 & 7.34E-05 & 0.0391 & 0.0528 \\
    \bottomrule
    \bottomrule
    \end{tabular}%
  \label{tab:case2-50-results}%
\end{sidewaystable*}%

\begin{sidewaystable*}
  \centering
  \caption{Modelling Results for case~II - 75\% Duty Cycle Interference}
    \begin{tabular}{ccccccccccccc}
    \toprule
    \toprule
    \multirow{2}[4]{*}{SINR (dB)} & \multirow{2}[4]{*}{$ P_{e} $} & \multicolumn{7}{c}{Transition probabilities}          & \multicolumn{2}{c}{MSE} & \multicolumn{2}{c}{$ \chi^2 $} \\
\cmidrule{3-9} \cmidrule(lr){10-11} \cmidrule(lr){12-13}         &       & $ a_{11} $ & $ a_{13} $ & $ a_{22} $ & $ a_{23} $ & $ a_{31} $ & $ a_{32} $ & $ a_{33} $ & IID   & Model & IID   & Model \\
    \midrule
    5.47  & 0.0496 & 0.9485 & 0.0515 & 0.9473 & 0.0527 & 0.3962 & 0.6038 & 0.0000 & 3.31E-04 & 3.19E-04 & 0.6539 & 0.2470 \\
    1.25  & 0.0495 & 0.9485 & 0.0515 & 0.9475 & 0.0525 & 0.3956 & 0.6044 & 0.0000 & 3.55E-04 & 3.44E-04 & 0.7741 & 0.7049 \\
    -2.58 & 0.0514 & 0.9464 & 0.0536 & 0.9454 & 0.0546 & 0.3885 & 0.6115 & 0.0000 & 4.26E-04 & 4.79E-04 & 0.9859 & 0.6809 \\
    -6.05 & 0.0526 & 0.9451 & 0.0549 & 0.9441 & 0.0559 & 0.3762 & 0.6231 & 0.0008 & 3.44E-04 & 2.99E-04 & 0.9822 & 0.5474 \\
    -10.68 & 0.0542 & 0.9445 & 0.0555 & 0.9420 & 0.0580 & 0.3654 & 0.6305 & 0.0041 & 3.22E-04 & 2.63E-04 & 0.5839 & 0.1270 \\
    -12.20 & 0.0654 & 0.9378 & 0.0622 & 0.9299 & 0.0701 & 0.3013 & 0.6630 & 0.0357 & 1.85E-04 & 1.55E-04 & 0.2404 & 0.0768 \\
    -14.08 & 0.0650 & 0.9418 & 0.0582 & 0.9290 & 0.0710 & 0.3050 & 0.6490 & 0.0460 & 1.23E-04 & 1.07E-04 & 0.1063 & 0.1003 \\
    -15.65 & 0.0772 & 0.9392 & 0.0608 & 0.9123 & 0.0877 & 0.2534 & 0.6825 & 0.0641 & 1.18E-04 & 1.12E-04 & 0.0638 & 0.0861 \\
    -16.31 & 0.0824 & 0.9312 & 0.0688 & 0.9093 & 0.0907 & 0.2330 & 0.7021 & 0.0649 & 1.15E-04 & 9.56E-05 & 0.0455 & 0.0359 \\
    -18.80 & 0.1013 & 0.9204 & 0.0796 & 0.8909 & 0.1091 & 0.1844 & 0.7155 & 0.1001 & 5.63E-05 & 4.28E-05 & 0.0284 & 0.0631 \\
    \bottomrule
    \bottomrule
    \end{tabular}%
  \label{tab:case2-75-results}%
\end{sidewaystable*}%


%

\begin{figure}[t]
	\centering
	\includegraphics[scale=0.4]{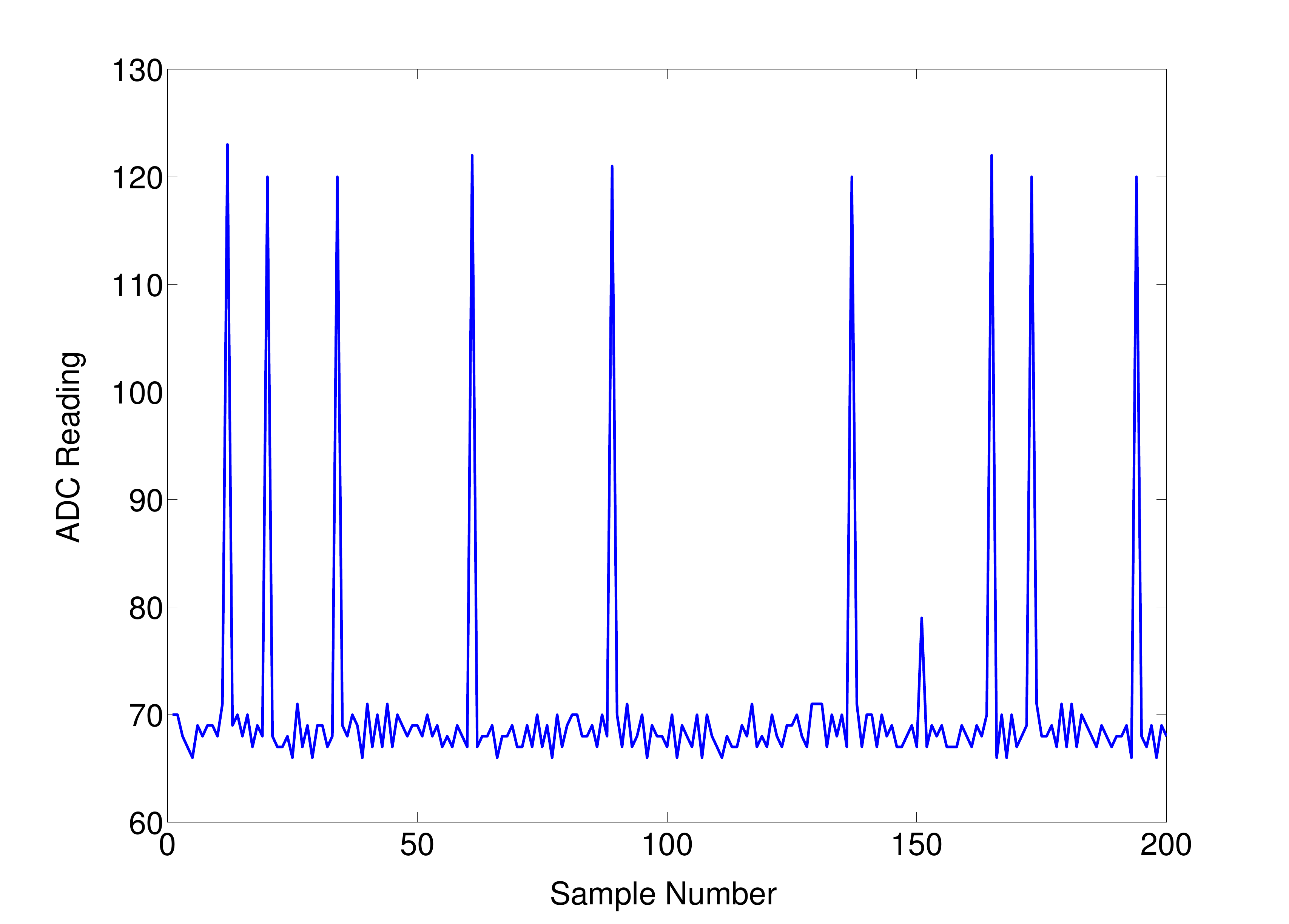}
	\caption{Samples of ADC readings used for SINR calculations.}
	\label{fig:adc-samples}
\end{figure}

\subsubsection{SINR Measurement Error}

The SINR values that were calculated and included as part of the results for case~II are lower than expected. This is as a result of limitations in the SINR measurements. SINR and SNR values were calculated using variance from ADC readings. Inaccuracies in the ADC readings led to variance values for the noise term being much being higher than anticipated. Figure \ref{fig:adc-samples} shows some of the ADC samples from the measurements used in case~II as part of the dataset for the calculation of the background noise power. Samples were supposed to be grouped into readings of the background noise only. However, because the grouping had to take place in the presence of the interfering PWM signal, some samples that were supposed to be put into the interference dataset were put into the noise dataset. These are evidenced by the spikes in ADC samples in Figure \ref{fig:adc-samples}. In other words, interference readings were unintentionally added to the noise readings. This invariably led to a higher denominator term the SINR calculation, thus, reducing the SINR values. This highlights the need for more robust methods in determining the interference levels from PWM signals.

\subsection{Log-likelihood}

For the generation of the models, the Baum-Welch algorithm was run 20 times. In order to test for the algorithm convergence, the log-likelihood ratios were plotted for cases~I and II, calculated using: 

\begin{equation*}
	\text{Pr}[\overline{O}|\Gamma] = \prod_{t=1}^{T}C_{t}
\end{equation*}

\begin{equation*}
	\text{log}_{10}\text{Pr}[\overline{O}|\Gamma] = \sum_{t=1}^{T}\text{log}_{10}C_{t}
\end{equation*}

\begin{figure}[!t]
\centering
\includegraphics[scale=0.38]{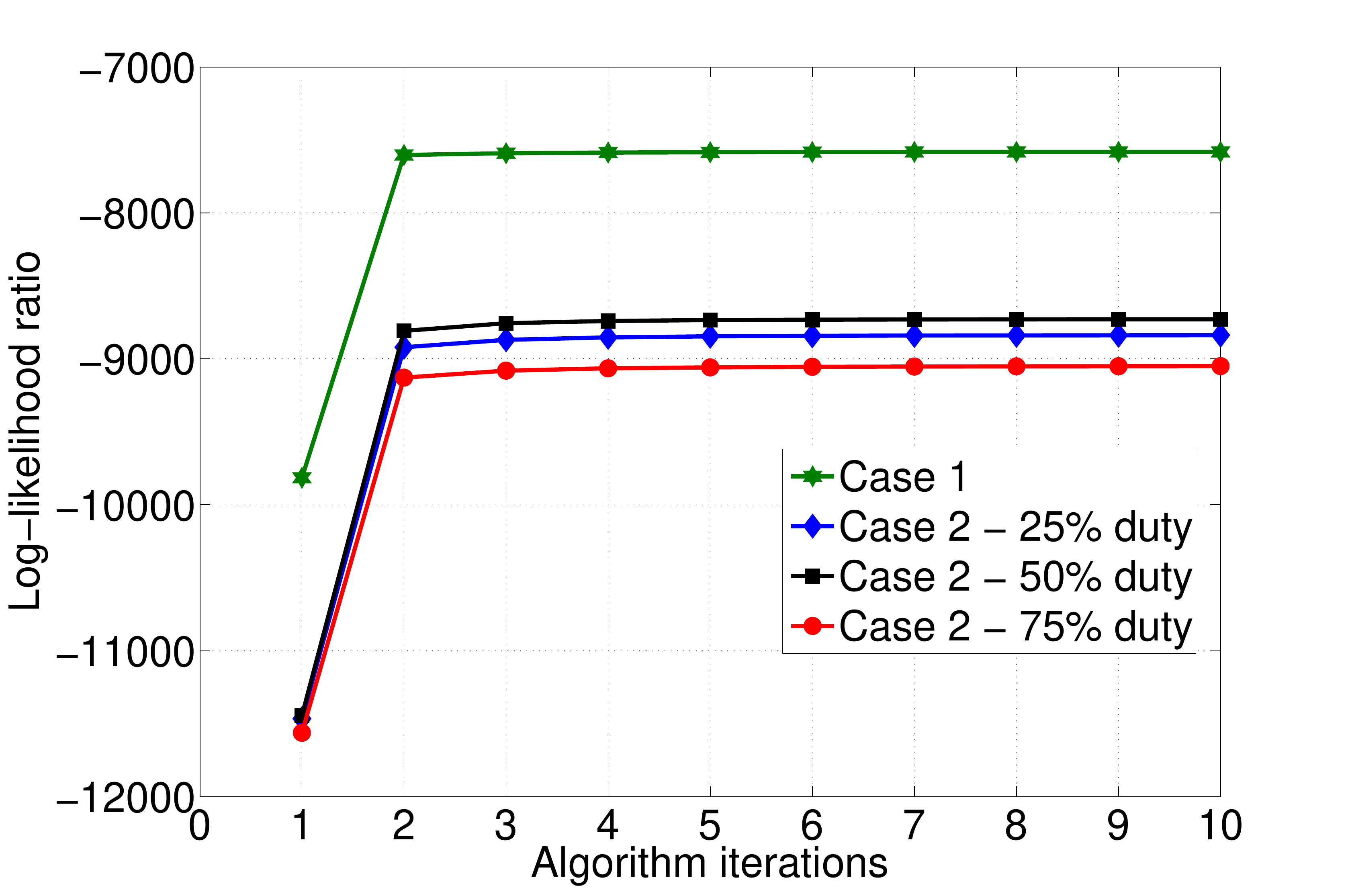}
\caption{Log-likelihoods for 10 iterations of the Baum-Welch algorithm, cases~I and II.}
\label{fig:log-likelihoods}
\end{figure}

The log-likelihood plots for four models are shown in Fig.~\ref{fig:log-likelihoods} for 10 iterations. Each of algorithms converged by the fifth iteration. The log-likelihood plots are also used to show how well the model fits the data the was used to train the models; in this case, it is the measured sequence. The closer the value is to 0, the better the fit. Fig.~\ref{fig:log-likelihoods} shows that the resulting model for case~I matched the measured sequence better than for case~II and their respective measured sequences.

\subsection{Discussion}

Error sequences generated from the models in case~I for different SNR values were all able to fit the measured error sequence. The models also provided a better fit compared to IID sequences. Results show that as the SNR values increase and $P_{e}$ increase for each test, the simulated error sequences (the models) are able to provide a more significant improvement in goodness of fit compared to the IID sequences. In other words, the models perform better in channels with less background noise. For example at the highest  SNR (7.68~dB), the difference between the IID and model $ \chi^2 $ values is 13.91; whereas, at the lowest SNR (4.28~dB) the difference is 0.122. This observation shows that as channel errors from background noise increase, the distribution of the errors becomes more uniform and less bursty.

\begin{figure}[!t]
\centering
\includegraphics[scale=0.4]{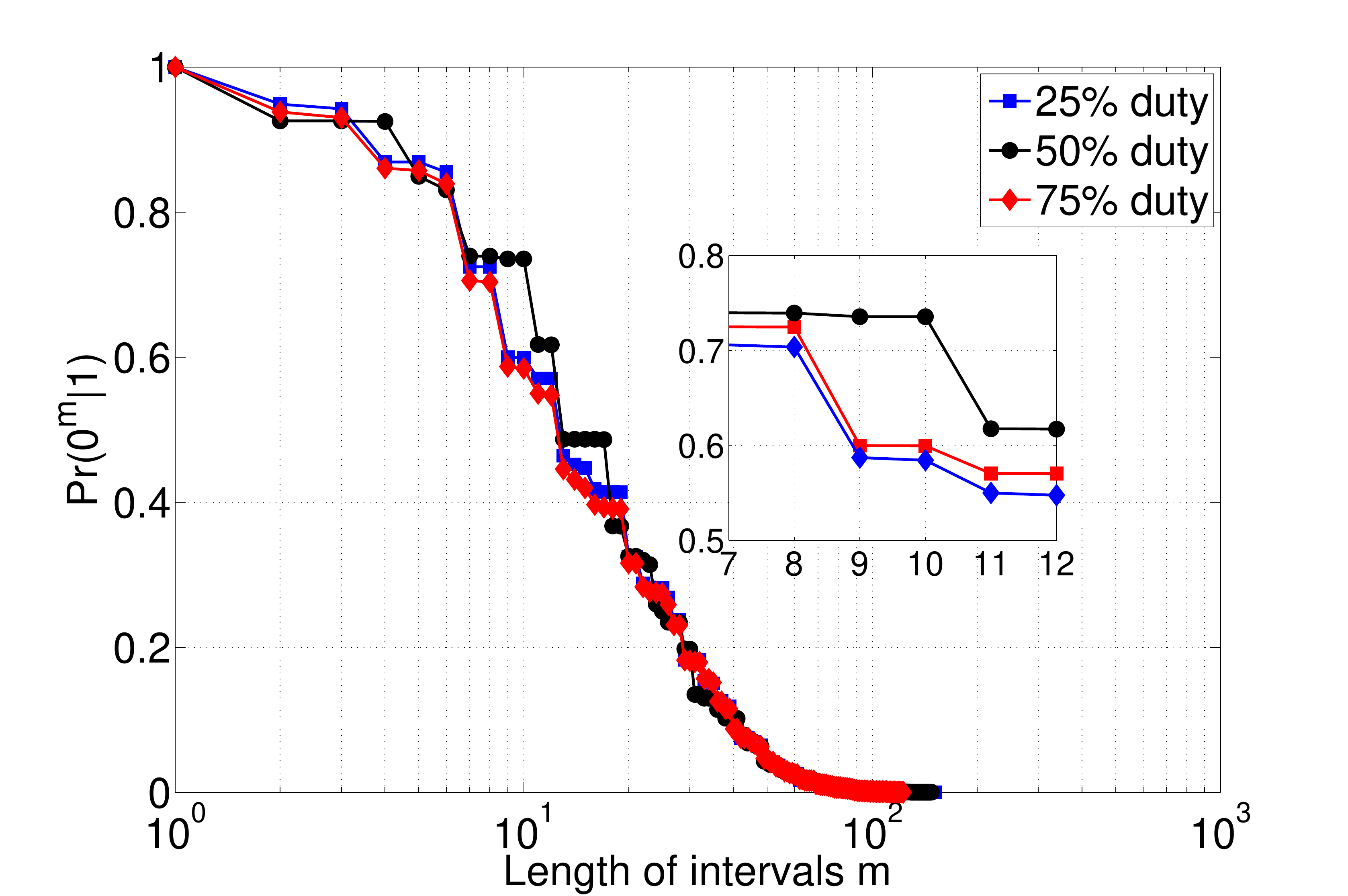}
\caption{Comparison of EFRD from the different PWM interference scenarios.}
\label{fig:duty-cycle-compare-zoom}
\end{figure}

The SINR values are significantly different for case~II compared to tests with similar $P_{e}$ values from case~I. This is due to the dominance of the interfering PWM signal which adds a $P_e$ increase of about 0.05. The errors from interference are also periodic in nature. This is due to the periodic interfering PWM signal. However, the periodic errors are not the same for different interference duty cycles. Fig. \ref{fig:duty-cycle-compare-zoom} shows the EFRD for the measured sequences for tests from each of the three duty cycles. The 25\% and 75\% duty cycle interference EFRD show very similar distributions. This is because the signals are 180$^\circ$ out of phase. Upon closer observation, Fig. \ref{fig:duty-cycle-compare-zoom} also shows evidence that the PWM transitions are also the source of errors, even with the optimal threshold detection. The gaps between these errors are dependant upon the gaps of the duty cycles. The PWM frequency is an order of magnitude higher than the transmission frequency. Thus, for a 50\% duty cycle interference, there is a higher probability of errors every 10 bits. On the other hand, for 25\% and 75\% there is a similar error probability, except it is every 8 bits. Thus, the channel with the 50\% duty cycle interference performs better that the 25\% and the 75\% scenarios.

These finding highlight the impact that PWM signals can have on low data rate VLC channels. It also justifies the need for error mitigation techniques. Future work can include developing models in a wider variety of scenarios, including non-line-of-sight and blockage conditions. There is also scope for research into mitigation techniques for co-existing VLC and smart lighting systems.

\section{Conclusion}\label{sec:conclusion}

Discrete channel models for different, low data rate indoor VLC scenarios have been presented, including different of types background noise and interference from other light sources. This interference included light from a nearby PWM modulated LEDs, such as those which would be part of an indoor smart lighting system at different PWM duty cycles. The models are based on Fritchman's SHMM and were developed by the Baum-Welch algorithm expectation maximisation using experimental data from OOK transmissions. Channel models are able to simulate binary error sequences with error distributions similar to the experimental error sequences and performed better than IID sequences. The models also highlighted the memory in the channel error sequences. In the presence of noise only, the models performed better at higher SNR values. These models also performed better than those without the presence of PWM interference. Yet, experiments did show that periodic errors are introduced from PWM interference. The 25\% and 75\% duty cycle interferences gave a similar error pattern, whilst the 50\% had the best performance. The models can be used for designing interference mitigation techniques for VLC smart lighting systems, error control codes and software simulations of indoor VLC channels with different types of interference.

\noindent\textbf{Funding:} This work was supported by the CSIR DST Inter-Programme Bursary (2017); the Wits University SITA bursary (2016-2017); and the Center for Telecommunications and Access Services funding at the School of Electrical and Information Engineering, Wits University (2016).

\section*{References}

\bibliography{bib}

\end{document}